\documentstyle[twocolumn,aps,epsfig]{revtex}
 \def\beq{\begin{equation}}
\def\eeq{\end{equation}}
\def\beqn{\begin{eqnarray}}
\def\eeqn{\end{eqnarray}}
\begin{document}

\title{Interpretation of Quantum Field Theories with a Minimal Length Scale}

\author{S.~Hossenfelder\thanks{email: sabine@physics.ucsb.edu}}

\address{Department of Physics, University of California\\ 
Santa Barbara, CA 93106-9530, USA}

\maketitle

\noindent
\begin{abstract} It has been proposed that the incorporation of an 
observer independent minimal length scale into the quantum field theories of the standard model
effectively describes phenomenological aspects of quantum gravity. The aim of this paper is to
interpret this description and its implications for scattering processes. 
\end{abstract}

\section{Introduction}
 
Quantum gravity is probably the most challenging and fascinating problem of physics in the 21st century. 
The most impressive indicator is the number of people working on it, even though so
far there is no experimental evidence that might guide us from mathematics to physical reality. 
During the last years, the priority in the field has undergone a shift towards the 
phenomenology and possible predictions \cite{Amelino-Camelia:2000ge,Amelino-Camelia:2000zs,Jacobson:2001tu,Amelino-Camelia:2002vw,Sarkar:2002mg,Konopka:2002tt,Alfaro:2002ya,Amelino-Camelia:2002vy,Heyman:2003hs,Jacobson:2003bn,Ahluwalia-Khalilova:2004dc,Smolin:2005cz}. The phenomenology of quantum gravity 
has been condensed into 
effective models which incorporate one of the most important and general features: a minimal 
invariant length scale that acts as a regulator in the ultraviolet. Such a minimal length scale 
leads to a generalized uncertainty relation and it requires a deformation of Lorentz-invariance 
which becomes important at high boost parameters. 

The construction of a quantum field theory that self-consistently allows such a minimal 
length makes it necessary 
to carefully retrace all steps of the standard quantization scheme. So far, there are various approaches 
how to construct a quantum field theory that incorporates a minimal length scale and the accompanying 
deformed special relativity ({\sc DSR}), generalized uncertainty principle ({\sc GUP}) and modified dispersion relation ({\sc MDR}). 
Most notably, there are approaches which start from the {\sc DSR} 
\cite{Magueijo:2001cr,Magueijo:2002am,Kimberly:2003hp,Daszkiewicz:2004xy,Girelli:2005dc,Konopka:2006fh}, 
the $\kappa$-Poincar\'e Hopf algebra 
\cite{Lukierski:1991pn,Majid:1994cy,Lukierski:1993wx,Kowalski-Glikman:2001gp,Bruno:2001mw,Kowalski-Glikman:2002we}
and those which start with the {\sc GUP} 
\cite{Hossenfelder:2003jz,Hossenfelder:2004up,Bhattacharyya:2004dy,Nozari:2005ex,Bhattacharyya:2005hq}. Besides this, there exists the possibility to
examine specific effects like reaction thresholds or radiation spectra starting from the {\sc MDR} without 
aiming to derive a full quantum theory in the first place \cite{Cavaglia:2004jw,Amelino-Camelia:2005ik,Camacho:2005qt}. 
Relations between several approaches have been investigated in \cite{Hossenfelder:2005ed}.

In this paper we aim to closely examine the ansatz starting with the GUP by paying special attention to the
interpretation of the effective theory. Since this starting point is conceptually different from the DSR-motivated one, 
it does not suffer from some of the problems that have been encountered within the latter, e.g. the conservation 
of momentum in particle interactions and the meaning of a highest energy scale for bound multi-particle states, the
so-called 'soccer-ball-problem'. As we will show, it is the treatment of
a single non-interacting particle which distinguishes both approaches.

This paper is organized as follows. In the next section we will investigate a picture of particle scattering
with additional strong gravitational interaction and motivate an effective model to extend the quantum field theories
of the standard model. In section three, some properties of the model are investigated. In section four 
it is examined in which cases the model can be applied with special
emphasis on the observer independence. In section five we analyze the relation to models starting with a deformation
of Lorentz transformations at high energies. We conclude in section six.

Throughout this paper we use the convention $c=\hbar=1$ and $G=1/m_{\rm p}^2$. Small Greek indices are
spacetime indices; small Latin indices label particle states.

\section{Motivation} 

A particle with energy close to the Planck mass, $m_{\rm p}$, is expected to significantly disturb space-time on
a distance scale comparable to its own Compton wavelength and thereby make effects of 
quantum gravity become important. A meaningful way to quantify how non-classical gravitational effects 
are is to examine the ability to describe spacetime as locally flat. The appropriate quantities are
the entries of the curvature tensor in a locally orthonormal basis, or, in case it is non-vanishing, the
curvature-scalar ${\cal R}$. Quantum effects should become strong, when 
${\cal R} m_{\rm p}^2 \sim 1$. However, provided that the rest mass of the particle 
itself is much smaller than the Planck mass, boosting the particle to high rapidity  
will not change the curvature, or the strength of quantum gravitational effects, it causes\footnote{From here on we assume that the 
rest mass of the particle is always much smaller than the Planck mass.}.  

Instead, to make the above statement precise, on has to refine its formulation: a concentration of energy 
high enough to cause strong curvature will result in significant quantum effects of gravity. Such a concentration
of energy might most intuitively be seen as an interaction process. In an
interaction process, the relevant energy is that in the center of mass (com) system, which we will denote with 
$\sqrt{s}$. Note that this is a meaningful concept only for a theory with more than one particle. 
However, it can also be used for a particle propagating in a background field consisting of many 
particles (like e.g. the {\sc CMB}). The scale at which effects of quantum gravity become important 
is when $\sqrt{s}/m_{\rm p} \sim 1$, for small impact parameters $\sqrt{t}\sim 1/b \sim m_{\rm p}$.

\begin{figure}[t]
\vskip 0mm
\vspace{0cm}
\centerline{
\psfig{figure=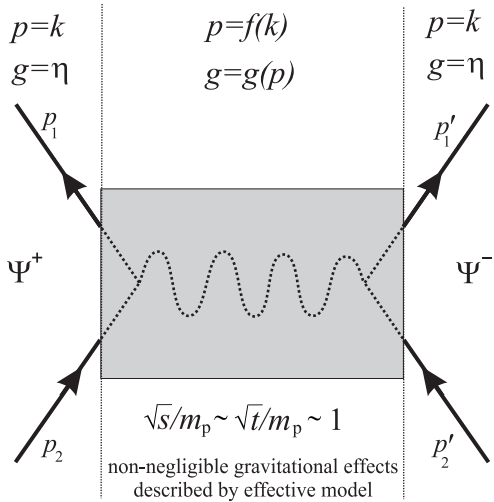,width=6.5cm}}
\vskip 2mm
\caption{In addition to the SM-interaction under investigation, strong gravitational effects accompany
the processes in the collision region. These effects are condensed in the effective QFT model with
a modified dispersion relation. Shown is the example of fermion scattering $f^+ f^- \to f^+ f^-$ 
($s$-channel). \label{fig1}}
\end{figure}

Let us consider the propagation of a particle with wave-vector $k_\mu$, when it comes into a spacetime 
region in which its presence will lead to a com energy close to the Planck scale. The concrete picture we 
want to draw is that of in- and outgoing point particles separated far enough and without noticeable 
gravitational interaction, that undergo a strong interaction in an intermediate region which we want to describe in an
effective way\footnote{An effective description as opposed to going beyond the theory of a 
point-like particle.}. We denote the asymptotically free in(out)-going states with $\Psi^+$ ($\Psi^-$), primes are
used for the momenta of the outgoing particles. This is schematically shown in Fig. \ref{fig1}.

In the central collision region, the curvature of spacetime is non-negligible and the scattering process
as described in the quantum field theories ({\sc QFT}s) of the standard model ({\sc SM}) is accompanied by 
gravitational interaction. We aim to find an effective description of this gravitational interaction,
which we expect to modify the propagator, not of the asymptotically free states, but of the particles that
transmit the interaction. The exchange particle has to propagate through a region with strong curvature, and
the  particle's propagation will dominantly be modified by the
energy the particle carries. 

Even on a classical level, the backreaction of a field propagating in space-time is involved. For simplicity, let us consider a massless scalar field $\phi$.
In principle, the evolution of space-time is described by Einstein's field equations
\beqn
R_{\mu\nu} - \frac{1}{2} g_{\mu\nu} {\cal R} = 8\pi G T_{\mu\nu}^\phi \quad, \label{efg}
\eeqn
where $G=1/m_{\rm p}^2$, and the source term is given by
\beqn
T_{\mu\nu}^\phi = \nabla_\nu \phi \nabla_\mu \phi - \frac{1}{2} g_{\mu\nu} g^{\kappa\epsilon} \nabla_\kappa \nabla_\epsilon \phi\quad.
\eeqn
The evolution of the field itself is the wave-equation in curved space
\beqn
g^{\kappa\epsilon}\nabla_\kappa \nabla_\epsilon \phi = 0\quad \label{waveeq}
\eeqn
which can be rewritten into partial derivatives
\beqn
g^{\kappa\epsilon} \partial_\kappa \partial_\epsilon \phi + g^{\kappa\epsilon} \Gamma^{\alpha}_{\;\;\kappa\epsilon} 
\partial_\alpha \phi= 0 \quad.
\eeqn
In the limit when the backreaction is small, the dependence of the metric on the field can be neglected. In this
case, one has $g_{\mu \nu} = g_{\mu\nu}(x)$. This leads to the formulation of a field theory in 
a possibly curved background.  When spacetime is asymptotically flat, such that 
$g_{\mu\nu}(x\to \infty) \to \eta_{\mu \nu}$, the equation of motion reduces to 
the familiar wave-equation which is solved by a superposition of modes of the form
\beqn
v_p \sim \exp \left( {\rm i}  \eta^{\mu\nu} p_\nu x_\nu  \right),
\eeqn
where $p$ fulfils the dispersion relation
\beqn
\eta^{\mu\nu} p_\mu p_\nu = 0 \quad.
\eeqn
However, in general the metric $g_{\mu\nu}$ will not only be a function of the space-time coordinates $x$, but also
a function of the derivatives $\nabla_\alpha \phi$, as dictated by Eqs.(\ref{efg}). The same is true for the Christoffelsymbols.
The general structure of Eq. (\ref{waveeq}) is then
\beqn
g^{\kappa\epsilon}(x,\nabla_\alpha \phi) \partial_\kappa \partial_\epsilon \phi + h^{\alpha}(x,\nabla_\alpha \phi) 
\partial_\alpha \phi= 0 \quad. \label{fullwave}
\eeqn
It is therefore natural to expect that in regimes where the gravitational interaction becomes important, 
the metric which the field propagates in will be a function of its energy (density).  

It  is most likely not possible to describe strong gravitational effects by using classical general relativity, and
the above motivation is not suitable to derive further details of the spacetime structure.
Instead, inherently new effects due to the quantum nature of spacetime will influence and eventually 
dominate the interaction processes. Such behavior has previously been investigated in various context using approaches from 
space-time foams, loop gravity or D-brane recoil 
\cite{Alfaro:2001rb,Urrutia:2002tr,Klinkhamer:2003ec,Ellis:2004ay,Livine:2004xy,Bojowald:2004bb}. These
investigations indicate that quantum effects result in a modified dispersion relation for the
propagating particles, 
which can also be formulated in terms of an energy dependent metric \cite{Kimberly:2003hp}.
  
The aim of the here discussed approach is to examine the additional gravitational 
interaction by means of an effective model\footnote{We do not consider
explicit production of real or virtual gravitons, or black hole formation.}. 
The {\sc QFT}s considered are modified in such a way that 
they capture one of the features that is generally expected to occur in quantum gravity 
\cite{Garay:1994en,Ng:2003jk,Hossenfelder:2004gj,Kempf:1998gk}: a minimal length scale. 
For this purpose, it is assumed that the exchange particles 
which mediate the SM interactions 
have to propagate through a region with a non-negligible quantum gravitational effects.
 
As motivated above and in \cite{Kimberly:2003hp} the assumption is  
that the quantum gravitational effects can be captured by the following description

\begin{itemize}
\item[1a] In a region of strong gravitational effects, the metric is dominated by the energy-dependence and
one has $g_{\mu\nu}=g_{\mu\nu}(\nabla_\alpha \phi)$ in $[-d,d]$. The coordinates are those of the 
asymptotically flat coordinate system with the interaction box in rest. \label{pt1}
\item[2a]  The wave-vector $k^\nu$ in the interaction region has
an upper bound $1/L_{\rm min}$. This is the fundamental assumption of a finite possible resolution. \label{pt2}
\end{itemize}

In a complete description, one would also expect the metric to be a function of coordinates: 
this behavior is simulated in the usual way by switching the interaction on and off in 
the central region, which allows us to remain in the momentum space description. 
In this case, the metric in the interaction region is not a function of the coordinates. Since 
the Christoffelsymbols are partial derivatives of the metric, covariant derivatives reduce to
partial derivatives and the wave-equation takes the form
\beqn
g^{\mu\nu}(\partial_\alpha \phi) \partial_\mu \partial_\nu \phi = 0 \quad, \label{waveint}
\eeqn
which is solved by a superposition of modes of the form
\beqn
u_k = \exp \left( {\rm i} \left( k_\nu x^\nu \right) \right), \label{mode}
\eeqn
where $k$ fulfils the dispersion relation
\beqn
g^{\mu\nu}({\rm i} k_\alpha) k_\mu k_\nu = 0 \quad. \label{mdr} 
\eeqn
Note that $k_\nu$ does not have a bound and that indeed both $k_\nu$ and $k^\nu$ are still completely normal vectors.
However, it is immediately apparent that under a transformation on $k_\nu$, the quantity $k^\nu = g^{\nu\kappa}(k) k_\kappa$
will transform non-linearly in $k$, see also \cite{Kimberly:2003hp}. To preserve parity, $g^{\mu\nu}$ should 
be an even function of $k_\alpha$, which also assures that no i's appear in the dispersion relation.

We will in the following refer to the dispersion relation as being a modified dispersion relation ({\sc MDR}) if 
\beqn
\eta^{\mu\nu} k_\mu k_\nu \neq 0 \quad.
\eeqn
Note, that this need not necessarily be the case for all equations of the form (\ref{mdr}). E.g. when the
energy dependent metric is of the form $g^{\mu \nu} = f(k) \eta^{\mu \nu}$ with some scaling function $f$, then
the dispersion relation (\ref{mdr}) implies the standard dispersion relation.

\section{Effective Description} 

Instead of dealing with an energy dependent metric as in Eq. (\ref{mdr}) 
the essence of the ansatz can also be captured by starting with the non-trivial relation between 
the globally conserved and the local quantities $p=f(k)$. This description has
been widely used in the {\sc DSR} literature, and in the following possible interpretations of this
approach are examined.

The functional form of the unknown relation $f(k)$ is where knowledge from an underlying theory 
has to enter. So far, the precise form of the function can not
be derived. However, the above mentioned general expectations allow us to constrain the 
form of the function. Such is that the Planck length acts
as a minimal length $L_{\rm min} \sim 1/m_{\rm p}$ in the sense that structures can not be 
resolved to smaller distances.  Note
again, that this statement is reasonable only for interaction processes since otherwise 'resolution' 
is not a meaningful concept.

It is nevertheless possible to construct a theory building up on single particles when one 
carefully keeps track of its meaning. E.g. the modified equation of motion Eq. (\ref{eom}) effectively
describes the gravitational interaction that the particle would undergo when it comes
close to a high com energy. The single particle meaning therefore is {a description of
what property a single particle would need to have in order to simulate the behavior of
quantum gravitational effects in the interactions}. To make this really clear: The
right way to describe the strong gravitational effects would be to include the appropriate 
quantized gravitational interaction, which is desirable
but so far an unsolved problem. Instead, we equip the point particle with an additional property ($k$ no
longer linear to $p$) that in interactions {\sl simulates effects that we expect from quantum gravity.}

According to the above, this relation between momentum $p$ and wave-vector $k$ to be such that no 
matter how high the energy of the particle gets, its wavelength can never become smaller than 
the minimal length. As a translation of 1a and 1b, The function $p=f(k)$ therefore has to fulfill the requirements:
\begin{itemize}
\item[1b] For energies much smaller than $1/L_{\rm min}$ the usual linear relation is found.
\item[2b] For large energies, $k$ asymptotically reaches $1/L_{\rm min}$.
\end{itemize}
And for a well defined relation we require
\begin{itemize}
\item[3] The function is invertible, i.e. it is monotonically increasing.
\end{itemize}

Theories of this type have been examined in various context as to their analytical structure
and phenomenological consequences \cite{Kempf:1994su,mlg,Maggiore:1993zu,Camacho}. The Lorentz-transformations
acting on the wave-vector in the collision region have to respect the above three points. This means
that for the quantity $k$ a deformed transformation is required which has an invariant minimal 
length $L_{\rm min}$ or an invariant maximal energy scale $m_{\rm p}$, respectively.

We write the relation between momentum and wave-vector in the form $p^\nu = f^\nu (k)$, which can be expanded in
a power series
\beqn
f^\nu(k) = \eta^{\nu\mu}\left( k_\mu + \sum_{l=1}^\infty \frac{{A^{(2l+1)}}_\mu^{\;\; \nu_1 \nu_2 ... \nu_{2l+1}}}{m_{\rm p}^{2l}} 
k_{\nu_1}k_{\nu_2}...k_{\nu_{2l+1}} \right) \nonumber 
\eeqn 
where it is taken into account that $p$ is odd in $k$. $A$ is a rank-$2l+1$-tensor with dimensionless
coefficients that, in accordance with the above point 2a, are constant with respect to space-time coordinates. 
Here,  $m_{\rm p}$ sets the scale for the higher order terms. Theories of this type have recently been
investigated in \cite{Magueijo:2006qd}.
 
The wave-vectors $k$ coincide with the momenta of the in- or outgoing particles far away from 
the interaction region, where space-time is approximately flat $g \sim \eta$. We will denote these 
asymptotic momenta by $p_i$. Putting the interaction into a box and forgetting 
about it, $\sum_i p_i$ is a conserved quantity\footnote{We assume that no
additional explicit losses e.g. in gravitons occur.}. The unitary operators of the Poincar\'e group act as
usual on the asymptotically free states. In particular, the whole box is invariant under translations 
$a_\nu$ and the translation operator has the form $\exp(-i a^\nu p_\nu)$ when applied to $\Psi^\pm$.  

In contrast to the asymptotic momenta $p$, the wave-vector $k$ of the particle in 
the interaction region will behave non-trivially because strong gravitational
effects disturb the propagation of the wave. In particular, it will not transform as a standard 
(flat space) Lorentz-vector, and obey the modified dispersion relation like Eq. (\ref{mdr}).  The action
of the Lorentz-group on states inside the interaction region will be modified and has been examined e.g.
in \cite{Magueijo:2001cr,Magueijo:2002am}. Though it is an important question to understand in which way the local 
gravitational interaction modifies operators of flat space {\sc QFT}, it is for our further investigation 
not necessary to deal with this issue\footnote{Note that an operator of the form 
$\exp(-i \tilde{a}^\nu k_\nu)$, where $k$ is not a Lorentz-vector leads to the conclusion that 
$\tilde{a}$ is not a Lorentz-vector either, and therefore requires some thought \cite{Kimberly:2003hp}.}.

Under quantization, the local quantity $k$ will be translated into a partial
derivate. 
One now wants to proceed from the single-$k$ mode (\ref{mode})
to a field and to the operator ${\hat k}_\nu = -{\rm i}\partial_\nu$. The corresponding
momentum-operator $\hat p$ should have the property
\beqn
\hat p^\nu v_k = p^\nu v_k = f^\nu(k) v_k \quad,
\eeqn
which is fulfilled by
\beqn
\hat p^\nu = f^\nu(- {\rm i} \partial) \quad,
\eeqn
since every derivation results in just another $k$. It is therefore convenient to define the 
higher order operator 
\beqn
\delta^\nu = {\rm i} f^\nu (- {\rm i \partial}) \quad. \label{delta}
\eeqn
Since $f$ is even in $k$, this operator's expansion 
has only real coefficients that are up to signs those of $f^\nu$. 
Note that $\delta^\nu$ commutes with $\partial_\kappa$.

From this one can further define the operator $\widetilde{\Box}$  which generates 
the wave-function Eq.(\ref{waveint})
\beqn
\widetilde{\Box} = g^{\mu\nu}(\partial_\alpha ) \partial_\mu \partial_\nu = \delta^\nu \partial_\nu \quad.
\eeqn
This modified D'Alembert operator plays the role of the propagator in the quantized theory. It captures
the distortion of the exchange-particles in the strongly disturbed background.

It is convenient to use the higher order operator $\delta^\nu$ in the setup of a field
theory, instead if having to deal with an explicit infinite sum.
Note, that this sum actually has to be infinite when the relation $p^\nu = f^\nu(k)$ has an 
asymptotic limit as one
would expect for an UV-regulator. Such an asymptotic behavior could never be achieved with a finite power-series.

The higher order operator $\delta^\nu$ fulfills the property (see Appendix B)
\beqn
\phi_\mu \left( \delta^\mu \psi \right) = - \left(\delta^\mu \phi_\mu \right) \psi + \mbox{total divergence.} \label{tbp}
\eeqn
This relation is essential for the usefulness of the operator as it allows to shift derivatives
in the derivation of the equations of motion from a variational principle. In particular,
the action for a scalar field\footnote{For a discussion of the Dirac-equation, gauge fields, and applications see e.g. 
\cite{Hossenfelder:2003jz,Hossenfelder:2004up}.} takes the form
\beqn
S &=& \int {\rm d}^4 x \sqrt{g }{\cal L} \quad,
\eeqn
with
\beqn
{\mathcal L} &=& \delta^\nu \phi~ \partial_\nu \phi \quad.
\eeqn
Using Eq.(\ref{tbp}), one then derives the equations from the usual variational principle to
\beqn
\delta^\nu \partial_\nu \phi = 0\quad. \label{eom}
\eeqn
The stress energy tensor is calculated in a similar way
\beqn
T_{\mu \nu}   
 &=& \partial_\nu \phi \partial_\mu \phi - \frac{1}{2}  g_{\mu\nu} {\mathcal{L}}\quad. \label{set}
\eeqn
This quantity is conserved with respect to $\delta^\nu$, i.e.
\beqn
\delta^\nu T_{\mu\nu} = 0 \quad. 
\eeqn
Note that it is not conserved with respect to $\partial^\nu = \eta^{\mu \nu}\partial_\mu$, since
one uses the equation of motion Eq. (\ref{eom}) for the conservation law. This becomes clear when one inserts a plane
wave. Since is was assumed that the dispersion relation is truly modified, it is $\eta^{\mu\nu}k_\nu k_\mu \neq 0$.
Instead, the relation needed for the stress-energy conservation is $g^{\mu \nu}k_\nu k_\mu = 0$. On the 
other hand, the quantity $T^{\mu\nu}$ is conserved with respect to $\partial_\nu$, but it will involve the 
energy-dependent metric from raising the indices in Eq.(\ref{set}).
  
The calculus with the higher order operator $\delta^\nu$ effectively summarizes the explicit 
dealing with the infinite series, as examined in \cite{Magueijo:2006qd}.

Normalized solutions to the wave-equation Eq. (\ref{eom}) can be found in the set of modes
\beqn
v_p(x) = \frac{1}{\sqrt{(2 \pi)^3 2 E}} {\exp} \left( {\rm i} k_\nu x^\nu \right) \label{pmode}\quad,
\eeqn
where $(E,p)=f(k_0,k)$. These modes solve the equation of motion
when $p$ fulfills the usual dispersion relation, or $k$ fulfills the {\sc MDR}, respectively.  
Therefore, the interpretation of $k$ in a geometrical meaning as a wave-vector is justified. 
When imposing boundary conditions, one sees that the relevant quantity is $k$ and not 
$p$ which makes it clear that modifications will arise whenever one attempts to confine 
the particle in a region of size comparable to the Planck length.  This, e.g. has 
consequences for the Casimir effect \cite{Harbach:2005yu,Bachmann:2005km} and for the 
evaporation of Planck-size black holes \cite{Cavaglia:2004jw,Amelino-Camelia:2005ik}.

According to the above discussion of the interaction region, we can now examine the properties of the
$S$-matrix. The quantity we measure for ingoing and outgoing states of a collision is typically not 
the wave-length of the particle but its ability to react with other particles. For 
scattering processes, the quantity $k$ therefore is a mere dummy-variable that justifies its existence 
as a useful interpretational device in intermediate steps, where it enters through the propagator defined
in Eq. (\ref{mdr}). It is in principle possible to calculate 
in $k$-space, however, eventually $k$ can be completely replaced by the physical momentum $p$. It
is important to note that from the construction of the model, $k$-space has finite boundaries, whereas the 
momentum-space is infinite with a squeezed measure at high energies that regulates the 
usually divergent integration (see e.g. \cite{Hossenfelder:2004up}).
 
In particular, the $S$-matrix is invariant under the standard Lorentz-transformation and conserves
the sum of in- and outgoing momenta. To see this, note that the unitary operator of the Poincar\'e 
group that belongs to a Lorentz boost $\Lambda$ and a translation $a$ acts on in- and
outgoing states in momentum space in the usual way
\beqn
U(\Lambda,a) \Psi^\pm (p) = \exp(-a^\mu (\Lambda p) _\mu) \Psi^\pm (\Lambda p) \quad. 
\eeqn
Therefore, the scattering matrix $S=\langle \Psi^-|\Psi^+ \rangle$ transforms according to\footnote{Further factors depend on the spin etc. 
and are not affected by $a$, see e.g. \cite{Weinberg}.} 
\beqn
&&S_{p_1,...,p_n,p'_1,...,p'_n} \sim S_{\Lambda p_1,...,\Lambda p_n,\Lambda p'_1,..., \Lambda p'_n} \times \nonumber \\ 
&& \exp\left[-i a^\mu \left( (\Lambda p_1)_\mu - ... - (\Lambda p_n)_\mu \right) \right] \nonumber \\
&& \exp\left[+i a^\mu \left( (\Lambda p'_1)_\mu  + ... + (\Lambda p'_n)_\mu \right) \right] \quad. \label{smatrix}
\eeqn
Since the left side is independent of $a$, so is the right side, which is possible only if
\beqn
(p_1)_\mu + ... + (p_n)_\mu - (p'_1)_\mu  - ...- (p'_n)_\mu  = 0  \quad,
\eeqn
which remains a true statement under Lorentz boosts. The total momentum of the in- and outgoing states
therefore is conserved in the standard way because these particles do not experience strong gravitational effects in
the asymptotic regions.  
 
\section{Observer independence of the Minimal Length}

The notion of a minimal length scale should be observer independent. At first sight, this seems to be in conflict with
the standard Lorentz-transformation since a boost would be able to contract a minimal length further. However, one has carefully to ask the right question. Consider two observers related by a standard 
boost, each having a ruler of minimal length in his rest frame. This is no contradiction as long as both do not
compare any quantity. The observer in the one system can not actually 'see' the length of the object in
the other system without probing it, which already involves an interaction process. 

Instead, one would ask both observers to perform the same experiment and measure which results arise from
initial conditions they have both agreed on. Such might e.g. be the ability to resolve smaller distances with
larger energy, the impossibility of which indicates the closeness of Planck-scale fuzziness and should be
equally impossible for all observers. Observer independence states that the outcome of such
experiments has to be the same in all rest frames. This reasonable expectation makes immediately 
clear what consequences a model like the here discussed can have and can not have. 

Consider a reaction made by one observer which
results in a cross-section $\sigma$ (Lorentz scalar) as a function of varying com energy $\sqrt{s}$ (Lorentz scalar). 
Increasing $\sqrt{s}$, at com energy close to the Planck scale, the reaction will stop probing 
smaller distances and the amplitude of the process will become (asymptotically) constant. In case the SM prediction
was an increasing function, the modified amplitude will be lowered in comparison, in case the 
SM prediction was a decreasing function, the modified amplitude will be raised in comparison.
One might say in general, the
amplitude stagnates. Since the cross-section
$\sigma({\sqrt s})$ takes into account the phase space of the outgoing particles, it is exponentially suppressed
at energies above the Planck scale \cite{Hossenfelder:2003jz}. 
One can expect collider signatures to be dominantly visible in the $s$-channel at large momentum transfer $t$.

However, if such a cross-section has a typical com energy $\sqrt{s_0}$ at which it has 
a sudden increase (crossing of reaction threshold), this threshold -- being a Lorentz scalar -- is
the same for both observers. In particular, a reaction threshold that has been observed in a laboratory
on earth to occur at a certain com energy, will in every inertial frame take place
at the same com energy unless observer independence is explicitly violated. 

Within the here discussed model, one expects a deviation of the cross-section relative to the {\sc SM} result, when 
the com energy for the reaction gets close to the Planck scale (or a new lowered fundamental scale). 
This however, is certainly not the case e.g. for the recently examined photopion production with a 
threshold of $\sim 1$ GeV that has been examined in earth's laboratories for decades. A modification of the
threshold for photopion production, which has been proposed to explain the non-observation 
of the {\sc GZK}-cutoff \cite{Amelino-Camelia:2000zs}, therefore is only possible when observer 
independence is violated. Indeed, one could use exactly this threshold to distinguish observers\footnote{One could
e.g. achieve such a scenario by allowing a 'scalar' to depend on the boost-parameter. By this, it becomes
possible to shift reaction thresholds. However, by doing so one has introduced a label to distinguish 
between observers through the value of that scalar.}. In the here 
presented approach, the threshold occurs at the same com energy in all reference frames, and thus the
{\sc GZK}-cutoff remains unmodified. This is due to the invariance of the cross-section under a boost from
the com frame of the earth experiment to that of the cosmic ray interaction for the same $\sqrt{s}$ (see also Appendix).
 
Nevertheless, it is possible to find modifications from phenomenological quantum gravity in processes 
where there is a natural candidate for a special reference frame. E.g. a long-distance propagation of particles through
the {\sc CMB} might reflect in a modified dispersion relation. In such a case, the particle constantly
propagates through an interaction region and therefore the {\sc MDR} applies. Even though the modification of the particle's 
propagation are tiny, they can add up over a long travel distance.  
An observable that has recently been investigated in this context
as to its possibility to reveal such quantum gravitational effects is the  time of flight or {\sc HBT} 
\cite{Camacho:2005qt} for $\gamma$-rays from
far away sources. In case the {\sc MDR} predicts a varying speed of light, the time of flight
can depend on the energy of the photon which could become detectable with {\sc GLAST} \cite{deAngelis:2000ji}.

Also, the meaning of quantum mechanics within the above introduced framework becomes
accessible using this interpretation. From a  non-linear relation $k(p)$, it follows that 
the uncertainty principle is generalized to
\begin{eqnarray}  
[\hat{x},\hat{p}]&=& + i \frac{\partial p}{\partial k} \quad
\Rightarrow\quad 
\Delta p \Delta x \geq \frac{1}{2}  \Bigg| \left\langle \frac{\partial p}{\partial k} 
\right\rangle \Bigg| \quad. 
\end{eqnarray}
In quantum mechanics, the interaction is not quantized but we describe a particle in a potential or
with boundary conditions respectively. Here, the potential plays the role of a background field
and sets the scale for the effects to become important. In such a scenario, the energy levels
of the hydrogen atom and the spectrum of the harmonic oscillator have been investigated 
\cite{hydrogen,Brau,Akhoury:2003kc,Hossenfelder:2003jz},  
and also the gyromagnetic moment (precision in a strong magnetic background field) of the 
muon \cite{Harbach:2003qz} receives corrections already at the quantum mechanical level.

Based on this, it is now possible to understand multi-particle bound states. Such a bound state of several
particles with small rest masses can eventually have a total mass higher than the Planck mass.
Assuming that the gravitational interaction of the bound particles is weak, quantum gravitational effects are negligible and
the system can be described within the standard {\sc QFT}. Most importantly, it will be boosted according to
the standard Lorentz-transformation like the free single particle (free from gravitational as well as SM interactions). 
There is no region of gravitationally strong interacting particles that
would require modifications. One might however probe such a bound system by using a high energetic 
beam in a Rutherford-like experiment. One would then again find a limit to the resolution of the
internal structure of the bound system. 

\section{Deformed Special Relativity}

From the above discussion it is now apparent that there are two conceptually different ways to include
a minimal length scale into the {\sc QFT}s of the {\sc SM} and to obtain an effective model. The model 
discussed here leaves the transformation of the free single particle unmodified since for such 
a particle there is no natural scale which could be responsible for quantum gravitational effects to become important. 

Starting from an effective description in terms of a modified dispersion relation in the 
interaction region, we have shown in how far this model is a reasonable candidate to describe 
strong gravitational effects in interactions. It has a clear interpretation 
for energy and momentum of the participating particles, and it does not suffer from the 
soccer-ball problem. As has been shown previously, the model has an ultraviolet regulator 
and  modified Feynman rules can be derived \cite{Hossenfelder:2004up}. This becomes possible 
by carefully asking
what observables we investigate. Interestingly, it has also been previously mentioned that the arising
problems in {\sc DSR} might be resolved by reconsidering the measurement process \cite{Liberati:2004ju}.

It is however also possible to start with a modification of the Lorentz-transformation for a 
free single particle and construct a {\sc QFT} based on this. For this, one assumes that the free
particle itself experiences the Planck mass as an upper bound on the energy scale\footnote{Since we have 
assumed that the rest mass of the particle is much smaller than the Planck mass, this can not
be the particle's own energy and leads back to the question of the measurement process.}.  

Such a {\sc DSR} is a non-linear representation of the Lorentz-group 
\cite{Magueijo:2001cr,Kowalski-Glikman:2002we,Lukierski:2002df}, which can be cast in a form similar
to the above used by introducing variables that transform under the usual representation of the
Lorentz-group, ${\cal P}=(\epsilon,\pi)$, and a general map $F$ to the variables that then obey 
the new deformed transformation law $P=F({\cal P})$.  However, due to the different interpretation 
of the free particle's quantities, the 'pseudo-variable' ${\cal P}$ is not the one identified with 
the physical four momentum. Instead, energy and momentum of the in- and outgoing particles are 
identified with those that have the deformed transformation behavior.   

As it has turned out over the 
last years, such attempts result in serious conceptual problems, the most important 
being the question of which quantities are conserved in interactions and the soccer-ball problem. 
Recently, important progress has been made as to how these models can be put on a solid 
base \cite{Magueijo:2006qd,Judes:2002bw,Girelli:2004ue} though the situation is not yet completely satisfactory and open
questions remain \cite{Liberati:2004ju,Toller:2003tz,Ahluwalia-Khalilova:2004dc}.

From the present day status, it is not possible to decide which is the right description of
nature. One might however lean on results from promising theories of quantum gravity and examine
e.g. the question in how far loop gravity \cite{Livine:2004xy,Bojowald:2004bb} or string theory have 
a minimal length. Though the appearance of a finite resolution in string-scattering has 
been examined \cite{Gross:1987ar,Konishi:1989wk,Amati,Yoneya:1989ai},
it remains to investigate in how far this is compatible with the {\sc DSR}-description of the point-particle 
limit of interaction processes, or whether a DSR can be accommodated in string theory \cite{Magueijo:2004vv}.

One should also keep in mind that these two mentioned approaches with a {\sc DSR} might not be the 
only possibilities to include a minimal length into {\sc QFT}. Another option might be to start with a 
modification of the interaction at smallest distances itself, and it is not a priori clear 
whether this can always be described in terms of the here discussed 
model\footnote{I thank Steve Giddings for bringing this into my attention.} or whether a more
general approach is necessary.
 
\section{Conclusions}

We have investigated an effective description of particle interactions in the presence of strong gravitational
effects. As a phenomenological description of quantum gravity, we have motivated the use of 
a modified dispersion relation, and we have interpreted the arising picture of the interaction process. 
Further, we have argued that a fundamental minimal resolution is an observer independent statement even though 
a free particle might still transform under standard Lorentz-transformations. We have shown that in this
case, the model has a clear interpretation for the conserved quantities and for the behavior of multi-particle
states. Based on this, we distinguished two conceptually different approaches towards a quantum field theory
with a minimal length, depending on the treatment of the non-interacting single particle.

\section*{Acknowledgments}

This work was supported by  the {\sc DFG} and by the Department of Energy under Contract DE-FG02-91ER40618. 
I thank Steve Giddings, 
Uli Harbach, Franz Hinterleitner, Stefan Hofmann, Tomasz Konopka, Jo\~ao Magueijo, J\"org Ruppert,
Stefan Scherer and Lee Smolin for helpful and stimulating discussions.

\begin{appendix}
\section{}

The threshold for a proton with four momemtum ${\bf p}=(E,\vec{p})$ and
a photon with four momentum ${\bf p}_\gamma=(\epsilon_\gamma, {\vec p}_\gamma)$ to produce a pion is given by the requirement that 
the total energy $s=({\bf p}+{\bf p}_\gamma)^2$ has at 
least to yield the rest masses of a produced pion $m_\pi$ and of the outgoing proton $m_{\rm Prot}$:
\beqn
s \geq (m_\pi + m_{\rm Prot})^2 \quad. \label{threshold}
\eeqn
A map like the above introduced $f$ or $F$ from quantities that transfrom under the usual Lorentz-tranformation
to those that obey the deformed transformation law, leaves a Lorentz-scalar a Lorentz-scalar (though these 
can differ by a factor which then necessarily is a constant, see also \cite{Judes:2002bw}). Applying
such a map to Eq.(\ref{threshold}) will therefore leave the inequality valid for all inertial systems. 

However, within the DSR-approach, the non-linear transformation law for the 'physical' momenta spoils the
Lorentz-invariance of this equation. Assuming that the single particle momenta transform according to the
new transformation, the additive quantity ${s}$ does not remain a Lorentz-scalar, which should be 
really carefully investigated as to whether it actually still allows observer independence.


\section{}

Let us start with examining an $n$th-order differential operator ${\mathcal D}^{(n)}$ of the form
\beqn
{\mathcal D}^{(n)} &=& b^{\nu_1} b^{\nu_2}  ...  b^{\nu_n} 
\partial_{\nu_1} \partial_{\nu_2} ... \partial_{\nu_n} \label{D}
\eeqn
Then one finds for some functions  $g$ and $h$ 
\beqn
\left( {\mathcal D}^{(n)} h \right) g &=& b^{\nu_1}  b^{\nu_2}  ...  b^{\nu_n} 
\left( \partial_{\nu_1} \partial_{\nu_2} ... \partial_{\nu_n} h \right) g \nonumber \\
&=& \partial_{\nu_1} \left( b^{\nu_1}  b^{\nu_2}  ...  b^{\nu_n}  
\left( \partial_{\nu_2} ... \partial_{\nu_n} h \right) g \right) \nonumber \\
&-& 
\left( b^{\nu_2} ... b^{\nu_n} \partial_{\nu_2} ... \partial_{\nu_n} h \right) 
\left( b^{\nu_1}   \partial_{\nu_1}  g \right)\quad\nonumber\\
&=& \mbox{t.d.} - 
\left( b^{\nu_2} ... b^{\nu_n} \partial_{\nu_2} ... \partial_{\nu_n} h \right) 
\left(b^{\nu_1}  \partial_{\nu_1}   g \right)~, \nonumber
\eeqn
where 't.d.' means 'total divergence' and is of the form $\eta^{\mu\nu}\partial_\mu A_\nu$. 
Repeat this step $n-1$ times to obtain
\beqn
\left( {\mathcal D}^{(n)} h \right) g &=&  (-1)^n h \left( {\mathcal D}^{(n)} 
g \right) +{\mbox{t.d.}} \quad. \label{1}
\eeqn
The operator $\delta^\mu$ is of the form
\beqn
\delta^\mu = \eta^{\mu \nu} \left( \partial_\nu + \sum_{l=1}^{\infty} \frac{\alpha^{(2l+1)}_\nu}{M^{2l}} 
{\mathcal D}^{(2l+1)} \right) \quad,
\eeqn
where the higher order contributions start with order $3$ in agreement with the above observation that the
series has only odd contributions.
It is
understood that the dimensionless coefficients $b$ in Eq.(\ref{D}) of the ${\mathcal D}^{(2l+1)}$-operators 
can depend on $l$
and can be translated in the $a$ coefficients of Eq.(\ref{delta}). 
Inserting into Eq.({\ref{tbp}}) gives
\beqn
\phi_\mu \left( \delta^\mu \psi \right) =  \eta^{\mu \nu} \phi_\mu
 \left( \partial_\nu \psi \right ) +  \eta^{\mu \nu} 
 \sum_{l=1}^{\infty} \frac{\alpha^{(2l+1)}_\nu}{M^{2l}} 
\phi_\mu \left( {\mathcal D}^{(2l+1)}   \psi  \right) \nonumber~.
\eeqn
The first part is the usual part,  
the second part can be rewritten with Eq.(\ref{1}) to
\beqn
\phi_\mu \left( \delta^\mu \psi \right) =  &-& \eta^{\mu \nu}  \left( \partial_\nu \phi_\mu \right) \psi \nonumber\\ 
&-&  
 \sum_{l=1}^{\infty}  \frac{\alpha^{(2l+1)}_\nu}{M^{2l}} 
\left( {\mathcal D}^{(2l+1)}   \phi_\mu \right) \psi   + \mbox{t.d.} 
\eeqn
and rearranging finally results in
Eq.(\ref{tbp}). It is worth noting that this does only work when only even powers of
the operators ${\mathcal D}^{(n)}$ appear.

\end{appendix}

\end{document}